# WHY THE SUNSPOT CYCLE IS DOUBLE PEAKED

Georgieva K.

Space and Solar-Terrestrial Research Institute-Bulgarian Academy of Sciences

Sofia, Bulgaria

*email: kgeorg@bas.bg*

**Abstract.** Many sunspot cycles are double peaked. In 1967 Gnevyshev suggested that actually all cycles have two peaks generated by different physical mechanisms, but sometimes the gap between them is too short for the maxima to be distinguished in indices of the total sunspot activity. Here we show that indeed all cycles have two peaks easily identified in sunspot activity in different latitudinal bands. We study the double peaks in the last 12 sunspot cycles and show that they are manifestation of the two surges of toroidal field – the one generated from the poloidal field advected all the way on the surface to the poles, down to the tachocline and equatorward to sunspot latitudes, and another one generated from the poloidal field diffused at midlatitudes from the surface to the tachocline and transformed there into toroidal field. The existence of these two surges of toroidal field is due to the relative magnitudes of the speed of the large-scale solar meridional circulation and the diffusivity in the solar convection zone which are estimated from geomagnetic data.

## 1. Introduction

By the term "solar activity", usually any type of variation in the appearance or energy output of the Sun is understood. Elements of solar activity are sunspots, solar flares, coronal mass ejections, coronal holes, total and spectral solar irradiance, etc. The most prominent evidence of solar activity and with the longest data record, though not geoeffective themselves but related to geoeffective active regions, are sunspots. Very big sunspots can be seen with naked eye, and old chronicles testify that they have been indeed observed even in ancient times. There is evidence that the Greeks knew of them at least by the 4[th] century BC, and the earliest records of sunspots observed by Chinese astronomers are from 28 BC. However, systematic observations of sunspots began only early in the 17[th] century after the telescope was invented.

As many other great discoveries, the sunspot cycle was discovered by chance. Heinrich Schwabe, a German pharmacist and amateur astronomer, was convinced that there must be a planet, tentatively called Vulcan, inside the orbit of Mercury. Because of the close proximity to the Sun, it would have been very difficult to observe Vulcan, and Schwabe believed one possibility to detect the planet might be to see it as a dark spot when passing in front of the Sun. For 17 years, from 1826 to 1843, on every clear day, Schwabe would scan the Sun and record its spots trying to detect Vulcan among them. He didn't find the planet but noticed the regular variation in the number of sunspots with a period of approximately 11 years, and published his findings in a short article entitled "Solar Observations during 1843" [1].

Apart from the observation that the number and area of sunspots increase and decrease cyclically every ~ 11 years, several other features of the sunspot cycle have to be taken into account in any theory trying to explain solar activity. The most important one is the finding of Hale published in 1908 [2] that sunspots are associated with strong magnetic fields, that sunspots tend to occur in pairs and that on one hemisphere the leading (with respect to the direction of the solar rotation) spots in all pairs have the same polarity, and the trailing spots in the pairs have the other polarity, while on the other hemisphere the polarities are oppositely oriented, and that in subsequent 11-yr sunspot cycles the polarities in the two hemispheres reverse [3]. Also, as the new solar cycle begins, sunspots first appear at higher heliolatitudes and as the cycle approaches minimum, their emergence zone moves equatorward – a rule known as Spörer's law but actually discovered by Spörer's contemporary Carrington [4].

Another empirical rule important for understanding the sunspot cycle, is Joy's law stating that the leading sunspot in a sunspot pair appears at lower heliolatitudes than the trailing sunspot, and the inclination of the sunspot pair relative to the equator increases with increasing latitude [5].

According to the solar dynamo theory developed by Parker in 1955 [6], the sunspot cycle is produced by an oscillation between toroidal and poloidal components, similar to the oscillation between kinetic and potential energies in a simple harmonic oscillator. The differential rotation at the base of the solar convection zone stretches the north-south field lines of the poloidal magnetic field predominant in sunspot minimum in east-west direction, thus creating the toroidal component of the field. The buoyant magnetic flux-tubes emerge piercing the solar surface in two spots with opposite polarities - sunspots.

This theory explains why sunspots appear in pairs, why the leading sunspots have the polarity of their respective poles, and why the polarities in the two hemispheres are opposite. To complete the cycle, the toroidal field must be next transformed back into poloidal field with the opposite magnetic polarity. Several classes of possible mechanisms are proposed to explain this process [7]. The most promising one recently is considered to be the so-called flux-transport dynamo mechanism first defined by Babcock in 1961 [8] and mathematically developed by Leighton in 1969 [9]. Due to the Coriolis force acting upon the emerging field tubes, according to Joy's law, the leading polarity sunspots are at lower heliolatitudes than the trailing polarity sunspots. Late in the solar cycle when sunspot pairs appear at very low heliolatitudes, the leading polarity sunspots diffuse across the equator and cancel with the leading polarity sunspots of the opposite hemisphere. The trailing polarity sunspots and the remaining sunspot pairs are carried toward the poles where the excess trailing polarity flux first cancels the flux of the old solar cycle and then accumulates to form the poloidal field of the new cycle with polarity opposite to the one in the preceding cycle.

Wang et al., 1991 [10] suggested that this flux-transport dynamo includes a large-scale meridional circulation in the solar convection zone which carries the remnants of sunspot pairs poleward at the surface. This circulation has been directly observed and confirmed from helioseismology (Hathaway, 1996, and the references therein, [11]; Makarov et al., 2001 [12]; Zhao and Kosovichev, 2004 [13]; Gonzalez Hernandez et al., 2006 [14]), magnetic butterfly diagram (Ivanov et al., 2002 [15]; Švanda et al., 2007 [16]), latitudinal drift of sunspots (Javaraiah and Ulrich, 2006 [17]). For mass conservation, the surface poleward circulation must be balanced by a deep counterflow at the base of the convection zone carrying the poloidal field back to low latitudes, transforming it on the way into toroidal field which emerges as the sunspots of the next cycle. This deep counterflow has not been yet observed, but has been estimated by the equatorward drift of the sunspot occurrence latitudes [18, 19]. Actually, this deep circulation carrying like a conveyor belt the flux equatorward explains the so-called Spörer's law – the equatorward motion of the sunspot occurrence zone.

Gnevyshev, 1963 [20] studied the evolution of the intensity of the solar coronal line at 5303 Å in different latitudinal bands during the 19th sunspot cycle, and found that there were actually two maxima in the 19th cycle: the first one, during which the coronal intensity increased and subsequently decreased simultaneously at all latitudes, appeared in 1957; the second maximum appeared in 1959-60 and was only observed at low latitudes, but below 15° it was even higher than the first maximum. Antalova and Gnevyshev, 1965 [21] checked whether this is a feature of the 19th cycle only, or of all cycles. They superposed the sunspot curves of all sunspot cycles from 1874 to 1962, and got the same result, that there are always two maxima in the sunspot cycle: the first one applies to all latitudes and appears simultaneously at all latitudes, and the second one occurs only at low latitudes. The relative amplitude of the two peaks and the time interval between them vary, so in some cycles they are seen as a single peak in latitudinal averages while in other cycles the gap between them known as "Gnevyshev gap" is clearly seen (Fig.1). Here and further, we use monthly values of the sunspot area smoothed by 13-point running averages with weight 0.5 for the first and last point and 1 for the other points [22].

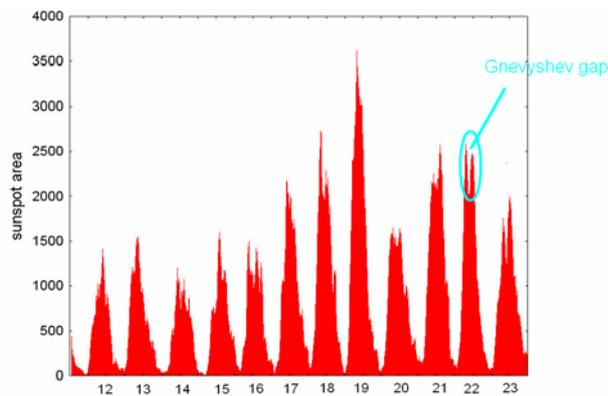

*Fig.1. Total global sunspot area with the Gnevyshev gap seen in some of the cycles*

For example, Norton and Gallagher, 2010 [23], studying the sunspot area and sunspot number summed over the whole northern and southern hemispheres and the over the whole disc, found Gnevyshev gaps in only 8 out of 12 cycles. However, when looking at different latitudinal bands, the Gnevyshev gap is clearly seen even in the cycles in which it is absent in the hemispheric or global totals. Fig.2 presents cycle 15 which according to [23] is single-peaked in the northern hemisphere and in the global sunspot area, but has two very distinct peaks appearing in different periods and in different latitudinal bands.

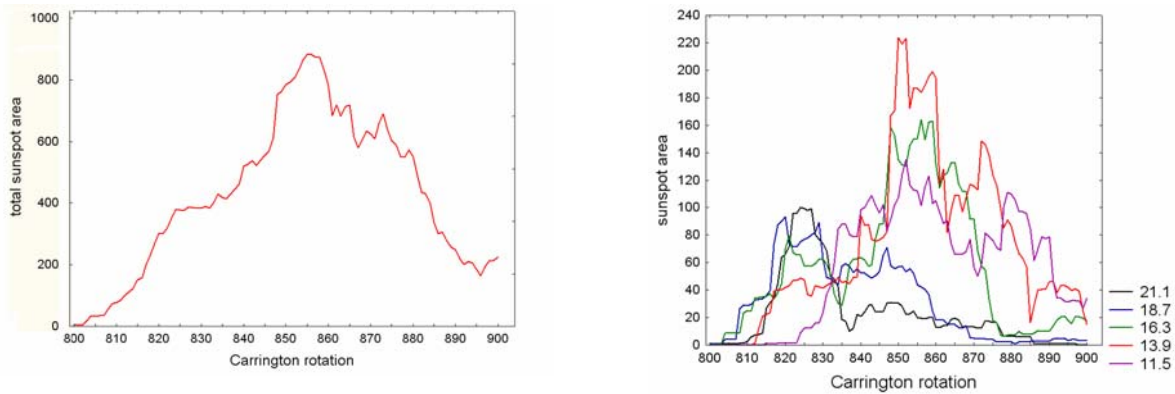

*Fig.2. (a) Total sunspot area in cycle 15; (b) sunspot area in separate latitudinal bins, averages over the two solar hemispheres.*

According to Gnevyshev, 1967 [24], the two maxima in sunspot activity result from different physical processes, and their existence means that the apparent gradual displacement of the sunspot occurrence zone to the equator is due to the superposition and changing relative importance of the two consecutive maxima. This conclusion, that the equatorward motion of the sunspot appearance zone is only apparent and is actually due to the superposition of the two different sunspot maxima, contradicts the explanation that it follows from the equatorward deep meridional circulation which is a critical part of the solar flux-transport dynamo theory.

The goal of the present study is to try to find an explanation of the double-peaked sunspot cycle in the framework of the flux-transport dynamo mechanism.

## 2. Diffusion and advection

The relation between the speed of the meridional circulation and the magnitude of the sunspot cycle is an indication of the regime of operation of the solar dynamo determined by the relative importance of advection (transport by the meridional circulation) and diffusion [25]. If advection is more important than diffusion in the upper part of the solar convection zone involved in poleward transport, a faster poleward circulation means less time for the leading polarity sunspots to cancel with their counterparts from the opposite hemisphere. On the way to the poles the leading and trailing polarity flux will cancel each other, and with less excess trailing polarity flux, less trailing polarity flux will reach uncanceled the poles to neutralize the polar field of the old cycle and to create the poloidal field of the new cycle. From this weaker poloidal field, weaker toroidal field will be generated, and the number of sunspots which are the manifestation of the toroidal field will be lower. In this case, there will be anticorrelation between the speed of the poleward meridional circulation and the amplitude of the following sunspot maximum.

If diffusion is more important than advection, the leading polarity flux will have enough time to cancel with the leading polarity flux of the opposite hemisphere, but the slower circulation will mean more time for diffusive decay of the flux during its transport to the pole, and hence a weaker poloidal field will result, respectively a weaker toroidal field of the next cycle. In this case, there will be correlation between the speed of the poleward meridional circulation and the amplitude of the following sunspot maximum.

If diffusion is more important than advection at the base of the solar convection zone where the toroidal field is generated, a faster equatorward circulation there will mean less time for diffusive decay of the flux during its transport through the convection zone, therefore stronger toroidal field and higher sunspot maximum of the next cycle. If advection is more important, higher speed will mean less time for generation of toroidal field, weaker toroidal field and lower sunspot maximum [26].

Unfortunately, both the speed of the solar meridional circulation and the diffusivity in the solar convection zone are largely unknown. Direct measurements of the surface meridional circulation are only available for less than three sunspot cycles [11, 27, 28], but its long-term variations and their correlation with the amplitude of the sunspot cycle are not known. The deep equatorward circulation has not been measured at all, its magnitude has only been estimated from the equatorward movement of the sunspot appearance zone [29]. The diffusivity in the upper part of the solar convection zone is estimated from the observed turbulent velocities and size of the convection cells, but its radial profile is completely unknown. Different authors have assumed different speeds of the surface and deep meridional circulation, and different values and radial distribution of the turbulent diffusivity, and based on the same flux-transport model, have obtained drastically different forecasts for the forthcoming solar cycle 24 [30].

We have proposed a method to evaluate the speed of the surface and deep meridional circulation from geomagnetic data [31]. Here we briefly explain this method and apply it to estimate the regime of operation of the solar dynamo and the origin of the two peaks of sunspot activity in the solar cycle.

## 3. Derivation of the solar meridional circulation and diffusivity

Even though we have no direct long-term observations of various solar processes, we do have means to reconstruct them. The Earth itself is sort of a probe registering solar variability, and records of the Earth's magnetic field contain evidence of Sun's activity effects.

The Earth's intrinsic magnetic field is basically dipolar, resembling the field of a bar magnet. Its magnitude is of order ~ 0.3 G (30,000 nT) at the Earth's surface near the magnetic equator, and twice that at the poles. At times of enhanced energy input as a result of the action of solar activity agents, the magnetic field is disturbed and its magnitude varies by about a percent of the main field due to currents in the ionosphere and magnetosphere.

Two types of solar activity agents are responsible for these geomagnetic disturbances, corresponding to the two faces of the sun's magnetism – the toroidal and poloidal components of the solar magnetic field. The coronal mass ejections – huge bubbles of plasma with embedded magnetic fields ejected from the solar corona, which like the number of sunspots are manifestation of the solar torroidal field, and the high speed solar wind streams, emanating from solar coronal holes, manifestation of the solar poloidal field. Consequently, geomagnetic activity has two peaks in the 11-year solar cycle (Fig.3).

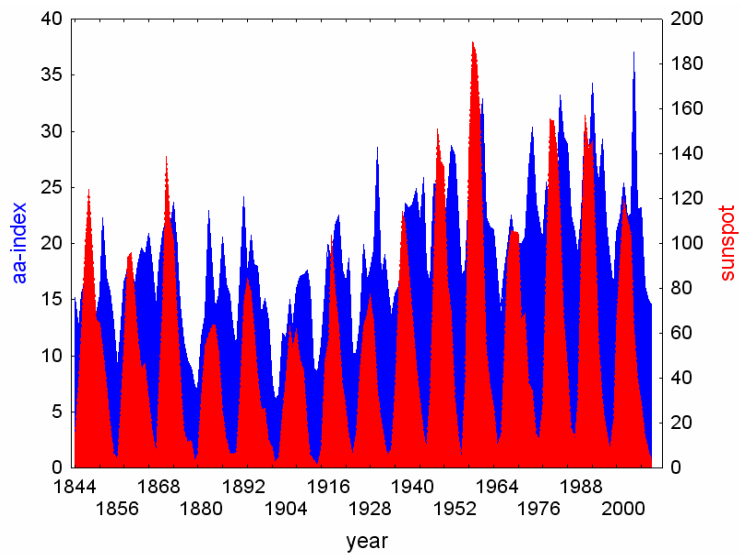

*Fig.3. Geomagnetic aa-index (blue) and sunspot number (red) since the beginning of the aa-index record.*

The first peak is due to solar coronal mass ejections which have maximum in number and intensity at sunspot maximum, and hence it coincides with the sunspot maximum. The second one is caused by high speed solar wind streams from solar coronal holes which have maximum on the sunspot declining phase [32, 33]. Like the peak in the number of sunspots, the first peak in geomagnetic activity can also be double or even multiple [34]. Fig.4 is an example of a cycle with a clearly visible double peak in sunspot activity reflected in geomagnetic activity (1989 and 1991), and a second peak in geomagnetic activity on the declining phase of the sunspot cycle (1994).

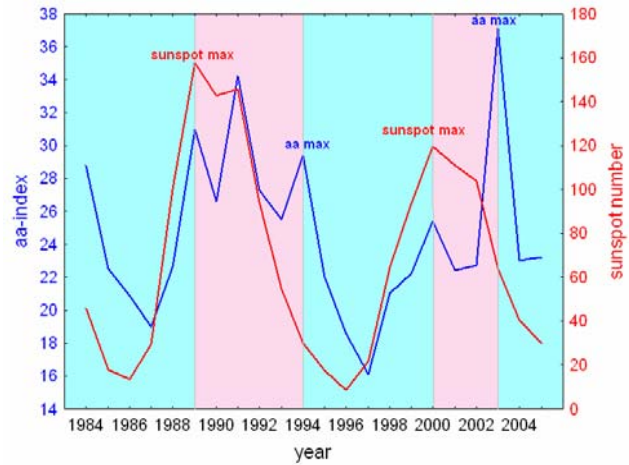

*Fig.4. Two geomagnetic activity maxima in the sunspot cycle used for the calculation of the speed of the surface poleward circulation: from the time between sunspot max and the following aa max (pink), and of the deep equatorward circulation from the time between aa max and the following sunspot max (cyan).*

What is the timing of the geomagnetic activity peak due to high speed solar wind from coronal holes? Coronal holes are observed at any time of the sunspot cycle, but their effect on the Earth varies. In sunspot minimum there are big polar coronal holes which however do not affect the Earth because the fast solar wind emanating from them does not reach the ecliptic (Fig.5a). In sunspot maximum there are small short-lived coronal holes scattered at all latitudes, giving rise to short and relatively weak high speed streams (Fig.5b). As shown by Wang et al., 2002 [35], geomagnetic activity reaches a maximum on the sunspot declining phase when polar coronal holes have already formed and low latitude holes begin attaching themselves to their equatorward extensions and growing in size, so the Earth is embedded in wide and long-lasting fast solar wind streams (Fig.5c).

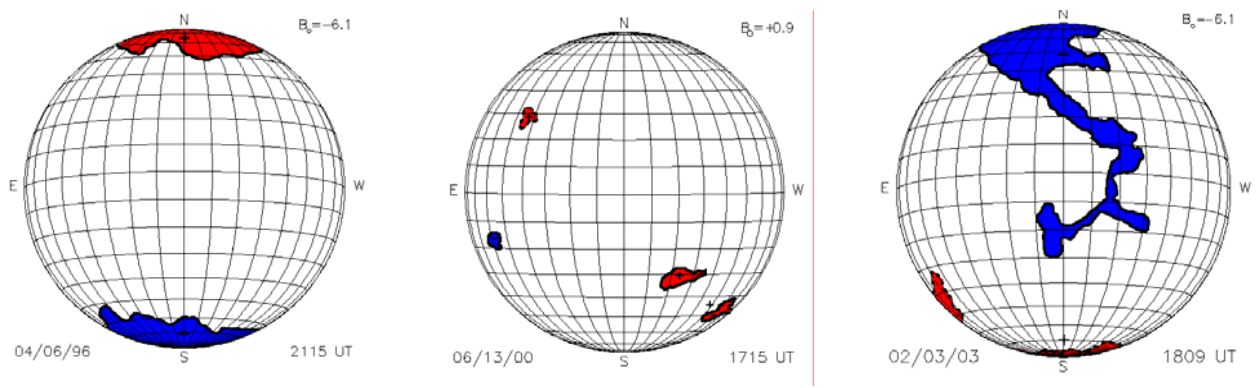

*Fig.5 (a) Sunspot min: large polar coronal holes; no coronal holes at low latitudes; (b) Sunspot max: small scattered short-living coronal hole at all latitudes; (c) Decline phase, when the trailing polarity flux reaches the poles. Coronal Holes data compiled by K. Harvey and F. Recely using NSO KPVT observations under a grant from the NSF.*

We therefore assume that the geomagnetic activity maximum on the declining phase of the sunspot cycle appears when the flux from sunspot latitudes has reached the poles. Hence, the time between sunspot maximum and geomagnetic activity maximum on the sunspot declining phase is the time it takes the solar surface meridional circulation to carry the remnants of sunspot pairs from sunspot latitudes to the poles, so from this time we can calculate the average speed of the surface poleward circulation Vsurf (Fig.6a).

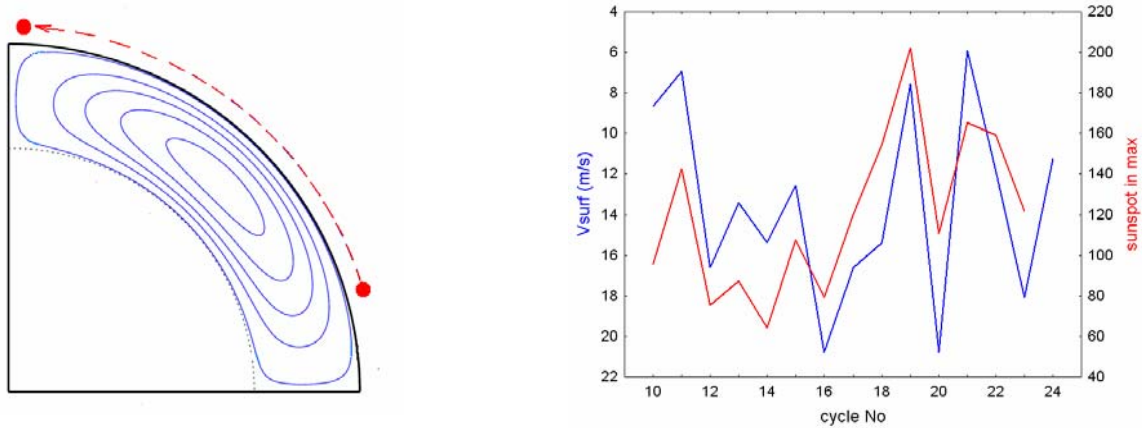

*Fig.6. (a) The distance traversed by the surface poleward circulation between sunspot maximum and the following geomagnetic activity maximum; (b) The calculated speed of the surface poleward meridional circulation Vsurf (blue line, note the reversed scale) and amplitude of the following sunspot cycle (red line).*

High and statistically significant anticorrelation is found between Vsurf and the amplitude of the following sunspot cycle: r=-0.7 with p=0.03 (Fig.6b, note the reversed scale of Vsurf). This means that advection is more important than diffusion in the upper part of the solar convection zone.

In order to check the reliability of the method, we compare the results to the available observational data. In the interval between sunspot cycle 10 and 23, Vsurf calculated by our method varies between 4 and 18 m/s averaged over latitude and over time. In the last cycle 23 for which direct observations are available, the calculated speed is 16 m/s and agrees remarkably well with results from helioseismology and magnetic butterfly diagrams which show latitude-dependent speed profile smoothly varying from 0 m/s at the equator to 20-25 m/s at midlatitudes to 0 m/s at the poles [11, 13].

From the time between the geomagnetic activity maximum on the sunspot declining phase and the following sunspot maximum (Fig.4), we can calculate the speed of the deep meridional circulation and/or the diffusivity in the bulk of the solar convection zone. What this time reflects depends on the diffusivity in the upper part of the convection zone. Three cases are possible according to the classification of Nitta and Yokoyama [35] and the estimations of Jiang et al., 2007 [36]: very low diffusivity (advection dominated regime), very high diffusivity (strongly diffusion dominated regime), and moderate diffusivity (moderately diffusion dominated regime).

**Advection dominated regime**

If the diffusivity in the upper part of the convection zone is very low which, according to Jiang et al., 2007 [36] means $\eta \sim 10^7$ m$^2$/s, the flux will make one full circle from sunspot latitudes to the poles, down to the tachocline and back to sunspot latitudes (Fig.7a). In this case, the time between the geomagnetic activity maximum on the sunspot declining phase and the next sunspot maximum is the time for the flux to sink to the base of the convection zone at polar latitudes, to be carried by the deep equatorward circulation to the equator, and to emerge as the sunspots of the new cycle. Assuming the speed of the downward transport of the flux equal to the speed of the deep equatorward circulation [36] and the time for the field tubes to emerge from the base of the convection zone to the surface equal to three months according to Fisher et al., 2000 [37], we can calculate the speed of the deep meridional circulation Vdeep (Fig.7b).

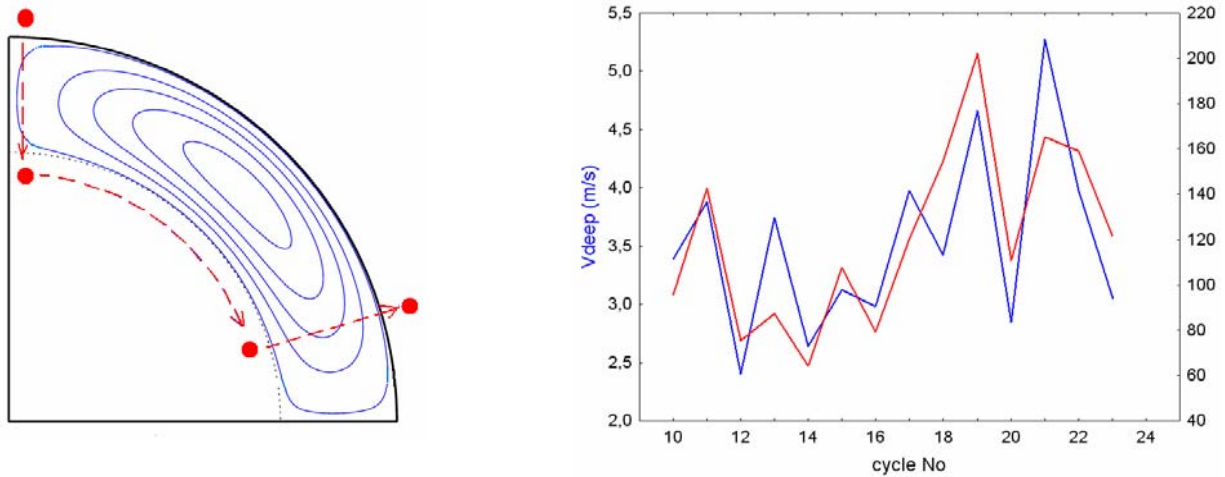

*Fig.7. (a) The distance traversed by the circulation between geomagnetic activity maximum and the following sunspot maximum in the case of very low diffusivity; (b) The calculated speed of the deep equatorward meridional circulation Vdeep (blue line) and amplitude of the following sunspot cycle (red line).*

The calculated Vdeep is between 2.5 and 5 m/s averaged over latitude and cycle, in excellent agreement with the estimations from the movement of the sunspot appearance zone [29] of speeds between 1.5 and 3 m/s at sunspot maximum, decreasing from high to low latitudes and from the beginning to the end of the cycle. This gives us further confidence in the reliability of the method.

The correlation between Vdeep and the amplitude of the following sunspot cycle is positive and highly statistically significant (r=0.79 with p<0.001). From the sign of this correlation we can evaluate the relative importance of diffusion and advection in the bottom part of the solar convection zone. The positive correlation obvious in Fig.7b means that diffusion is more important than advection there, so faster circulation means less time for diffusive decay of the flux during its transport through the convection zone, therefore stronger toroidal field and higher sunspot maximum of the next cycle.

**Strongly diffusion dominated regime**

The other extreme – very high diffusivity and strongly diffusion-dominated regime in the upper part of the convection zone – occurs when $\eta \sim 2\text{-}9 \cdot 10^8$ m$^2$/s and $\eta/u_0 > 2 \cdot 10^7$ m where $u_0$ is the maximum surface circulation speed [38]. In this case the time from the geomagnetic activity maximum on the sunspot declining phase to the next sunspot maximum will be the time it takes the flux to diffuse through the convection zone $T=L^2/\eta$ where L is the thickness of the convection zone (Fig.8a), and from it we can calculate the average diffusivity in the bulk of the convection zone and the ratio $\eta/u_0$ (Fig.8b). This is the upper limit of diffusivity because it is calculated under the assumption that all of the flux diffuses directly to the tachocline before it can reach the poles.

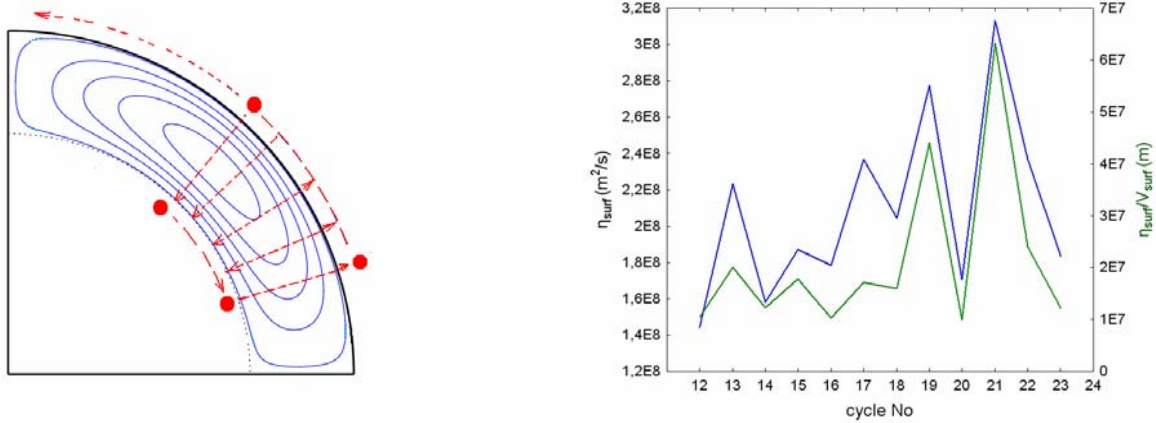

*Fig.8. (a) Diffusion through the convection zone in the case of very high diffusivity in the upper part of the convection zone. (b) The calculated diffusivity in the upper part of the solar convection zone η (blue line) and ratio of the diffusivity to the maximum surface circulation η/u$_0$ (green line).*

The values presented in Fig.8b are supported by estimations of the diffusivity in the upper part of the convection zone based on the observed turbulent velocities and size of convection cells [9, 36], and considerations about the correlation between the two solar hemispheres, and between the strength of the polar field and the amplitude of the following sunspot maximum [39].

## 4. The sunspot cycle in moderately diffusion-dominated regime

As seen from Fig.8b, even under the assumption of strong diffusion so that that all of the flux diffuses through the convection zone before reaching the poles, both calculated $\eta$ (~ 1-2.10$^8$) and $\eta/u_0$ (~5.10$^6$-2.10$^7$) are not high enough for strongly diffusion-dominated regime, and on the other hand - not low enough for fully advection-dominated regime. According to Jiang et al. [36], in the case of intermediate diffusivity $\eta \sim$ 1-2.10$^8$ m$^2$/s ("moderately diffusion-dominated regime"), a part of the flux short-circuits the meridional circulation, another part makes a full circle to the poles, down to the base of the convection zone and equatorward to sunspot latitudes (Fig.9).

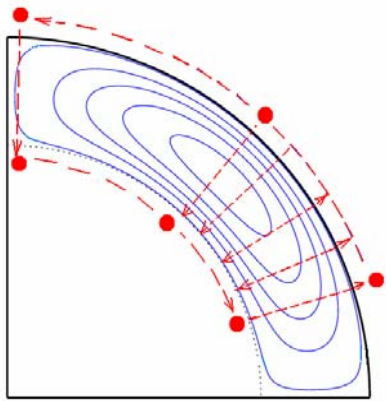

*Fig.9. Moderately diffusion-dominated regime: a part of the flux diffuses through the convection zone, "short-circuiting" the meridional circulation, another part makes a full circle to the poles, down to the base of the convection zone and equatorward to sunspot latitudes.*

If the solar dynamo operates in moderately diffusion-dominated regime in the upper part of the convection zone, the sunspot cycle will be a superposition of two surges of toroidal field: one generated from the poloidal field diffused across the convection zone, and another one - from the poloidal field advected by the meridional circulation. To check this we have plotted the sunspot area as a function of time and latitude (from http://solarscience.msfc.nasa.gov/greenwch.shtml). The sunspot area is given in 50 latitude bins (25 in each hemisphere) distributed uniformly in Sine(latitude) for each sunspot cycle from 12 to 23. Here we have used the averages over the two hemispheres in the respective bins.

Fig.10, (a) and (b) demonstrates the two peaks in cycle 16. The first one is centered at Carrington rotation 965 and appears simultaneously in a wide latitudinal range between 26.1 and 18.7° heliolatitude, the second one appears later and moves from 16.3° in Carrington rotation 981, to 13.9° in rotation 1003, to 9.2° in rotation 1018, and to 4.6° in rotation 1024. We identify the first peak with the flux diffused across the convection zone, and the second one – with the flux advected all the way to the poles, down to the tachocline and back equatorward to sunspot latitudes. Fig.10 (c) presents a surface plot of the sunspot area in cycle 16 (averaged over the two hemispheres) as a function of time and latitude. The cyan lines delineate the evolution of the two surges of sunspot activity, with the vertical lines indicating the diffusive generated peak occurring simultaneously in a wide latitudinal interval, and the tilted lines indicating the advection dominated peak progressing equatorward.

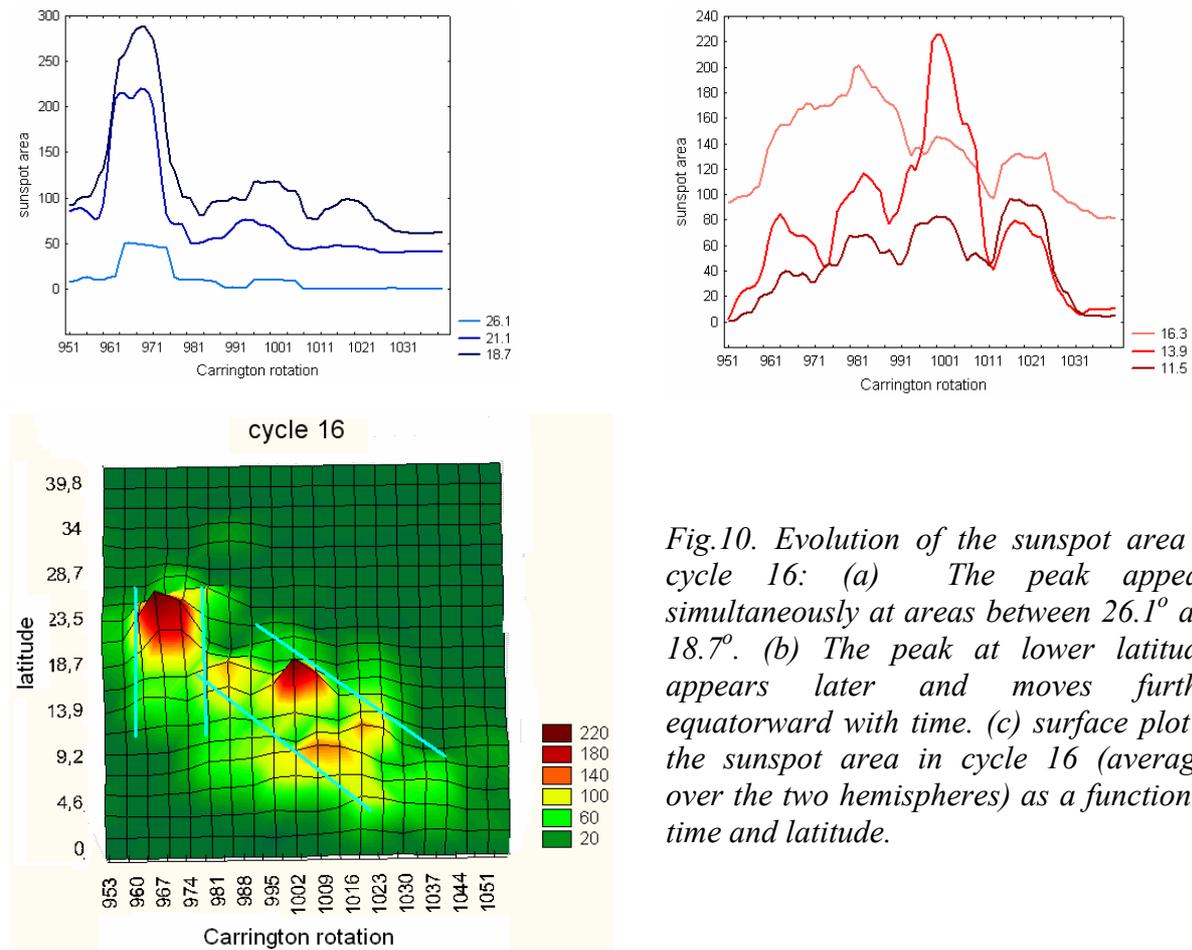

*Fig.10. Evolution of the sunspot area in cycle 16: (a) The peak appears simultaneously at areas between 26.1° and 18.7°. (b) The peak at lower latitudes appears later and moves further equatorward with time. (c) surface plot of the sunspot area in cycle 16 (averaged over the two hemispheres) as a function of time and latitude.*

The diffusion generated peak seems to appear earlier and at higher heliolatitudes in all cycles from 15 to 19. The order is reversed in cycles 12-14 and 20-23: first the advection generated peak at higher latitudes, then the diffusion generated peak at lower latitudes. An example (cycle 21) is shown in Fig.11. First the advection generated peak appears at Carrington rotation 1650 and moves to rotation 1690 from about 20$^o$ to about 10$^o$; the second peak, centered around Carrington rotation 1705, appears simultaneously at all latitudes below 15$^o$

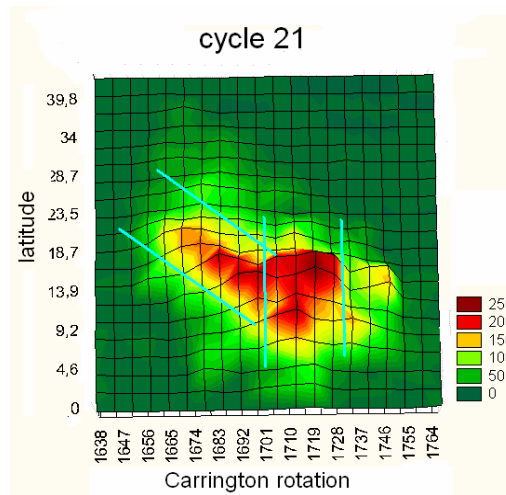

*Fig.11. The same as Fig.10c for cycle 21*

Fig.12 presents the surface plots of the sunspot area in all cycles from 12 to 23 (averaged over the two hemispheres) as a function of time and latitude. It seems that the order changes either in ascending and descending phases of the secular cycle or in consecutive secular cycles (Fig.13). At present it is difficult to understand the reason for this, but it obviously has a connection with the long-term variations in solar activity and can give additional information about solar dynamo. Note that the colour codes are different in the plots for the different cycles.

**Summary and conclusion**

Solar dynamo can operate in different regimes determined by the relative importance of diffusion and advection in the upper and bottom parts of the solar convection zone. Based on estimations for the speed of the surface poleward meridional circulation and diffusion in the upper part of the convection zone, we have demonstrated that the dynamo operates there in moderately diffusion dominated regime, in which a part of the flux short-circuits the meridional circulation and diffuses directly to the bottom of the convection zone at midlatitudes, another part makes a full circle to the poles, down to the base of the convection zone and equatorward to sunspot latitudes. These two parts of the flux, when transformed by the differential rotation at the base of the convection zone, give rise to two peaks of sunspot activity which are close but not exactly coinciding. In this way the double peaked sunspot cycle and the Gnevyshev gap have their natural explanation in the flux transport dynamo theory.

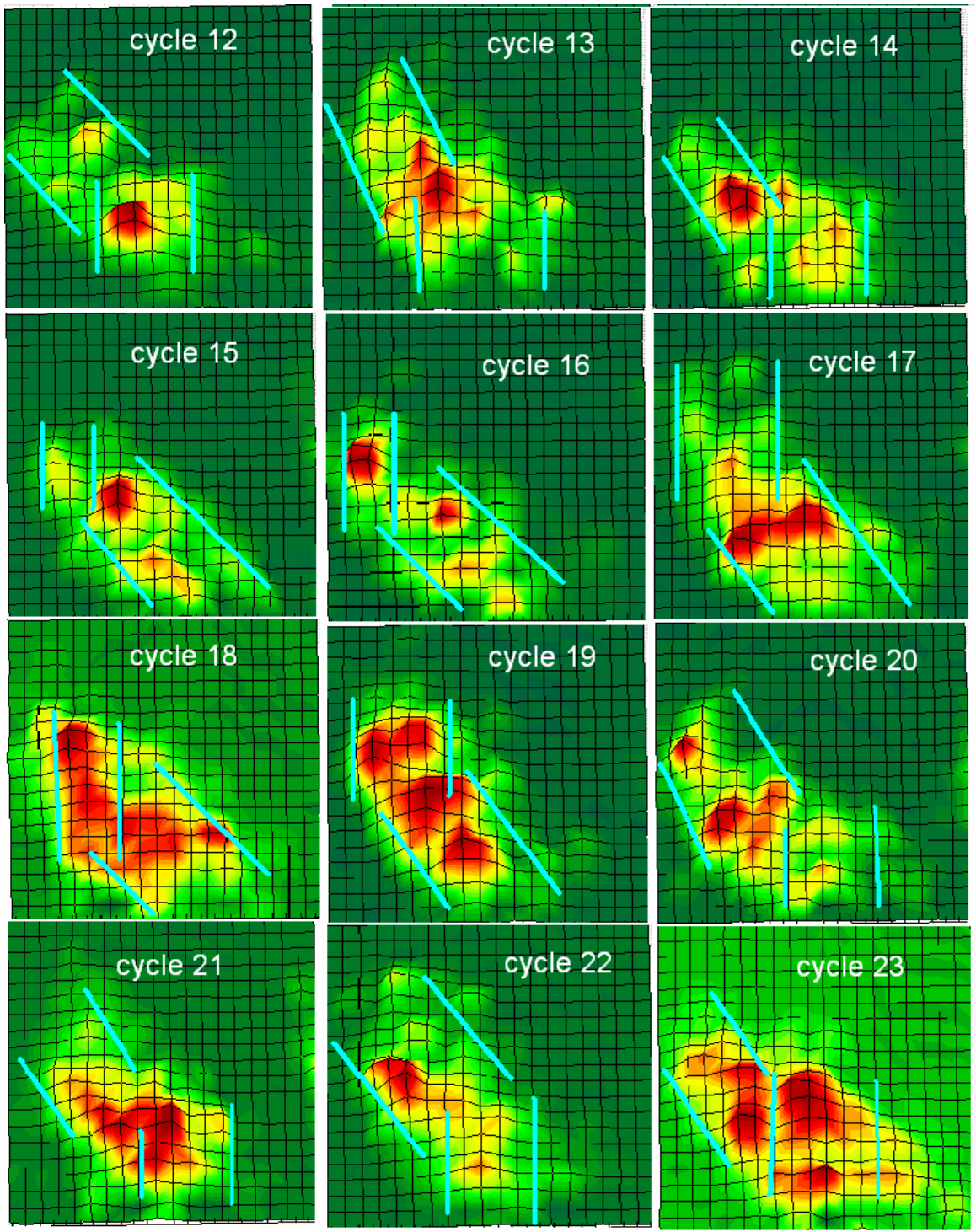

*Fig.12. The same as Fig.10c for cycles from 12 to 23. Note that the colour code is different for the different plots.*

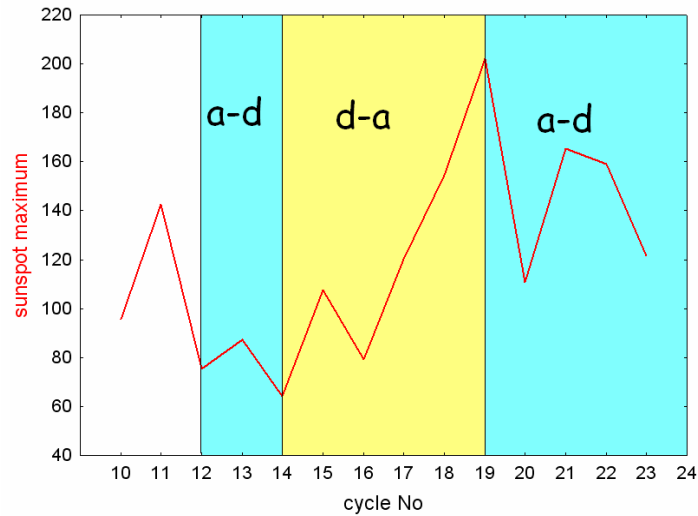

*Fig.13. Advection-dominated before diffusion-dominated peak ("a-d") and diffusion-dominated before advection-dominated peak ("d-a") in the secular solar cycle.*